\documentclass[dvips]{amsart}
\usepackage{amsmath}%
\usepackage{amsfonts}%
\usepackage{amssymb}%
\usepackage{graphicx}
\theoremstyle{plain}
\newtheorem{acknowledgment}{Acknowledgement}

\numberwithin{equation}{section}
\begin{document}
\title[Coulomb gauge]{Convergence of Feynman integrals  in Coulomb Gauge QCD}
\author[Andrasi]{A. Andra\v si}
\curraddr[Andrasi]{Vla\v ska 58, Zagreb, Croatia}

\email[Andrasi]{aandrasi@irb.hr}

\author[Taylor]{J. C. Taylor}
\curraddr[Taylor]{DAMTP, University of Cambridge, Cambridge, UK}
\email[Taylor]{jct@damtp.cam.ac.uk}

\date{June 27, 2014}

\begin{abstract}
At 2-loop order, Feynman integrals in the Coulomb gauge are divergent  over the internal energy variables.
Nevertheless, it is known how to calculate the effective action, provided that the external gluon fields are all transverse.
We show that, for the two-gluon Greens function as an example, the method can be extended to include longitudinal external fields.
The longitudinal Greens functions appear in the BRST identities.
As an intermediate step, we use a flow gauge, which interpolates between the Feynman and Coulomb gauges.

\end{abstract}

\maketitle
\clearpage

\section{Introduction}
The Coulomb gauge in QED and QCD has some theoretical and practical attractions. It is the only well-defined (or apparently well-defined)
gauge which is explicitly unitary, with no propagating ghosts. It has been used in lattice calculations, for example\cite{lattice}, and in investigations of confinement \cite{confinement}.
(We concern ourselves with perturbation theory, so the Gribov horizon is irrelevant \cite{gribov}.)

But the gauge is beset with \textit{energy divergences}, that is Feynman integrals which are divergent over the internal energy
variables ($p_0,q_0,r_0$ in our notation). We use dimensional regularization for ultra-violet divergences, but this
cannot control energy divergences. Firstly, there are linear energy divergences, but it is relatively easy to combine
graphs so that these cancel \cite{lineardivs}. Better, they are removed by using the phase-space formalism with
first order equations of motion, and we use this throughout the present paper.

There is a more subtle type of energy divergence in 2-loop graphs and sub-graphs, in the shape of the integral (we denote Lorentz vectors as
$p=(p_0,\textbf{P})=(p_0,P_i)$ etc)
\begin{equation}
\int dp_0 dq_0 \frac{p_0}{p_0^2-\textbf{P}^2+i\epsilon}\frac{q_0}{q_0^2-\textbf{Q}^2+i\epsilon}.
\end{equation}
This difficulty has been dealt with in two different ways. First, it was recognized as being connected with the operator ordering
of the Coulomb Hamiltonian \cite{christlee} (see also \cite{chengtsai}), and the correct ordering required the addition to the Hamiltonian of two  terms of 2-loop order called
$V_1$ and $V_2$. At the same time, to avoid double counting, the Feynman rules have to be specified  in such a way that
the integral (1.1) is defined to be zero (which, however, contradicts the identity (1.3) below).

The second approach to the integral (1.1) \cite{doust}  was to show systematically that graphs can be combined so
that only the convergent combination ($i\epsilon$s are understood and $p+q+r=k$ the external energy-momentum)
\begin{equation}
\Xi =
\left[\frac{p_0}{p_0^2-\textbf{P}^2}\frac{q_0}{q_0^2-\textbf{Q}^2}+\frac{q_0}{q_0^2-\textbf{Q}^2}\frac{r_0}{r_0^2-\textbf{R}^2}
+\frac{r_0}{r_0^2-\textbf{R}^2}\frac{p_0}{p_0^2-\textbf{P}^2}\right]
\end{equation}
occurs. Then the energy integrals are convergent, and give a result independent of the external energy $k_0$:
\begin{equation}
\int dp_0 dq_0 dr_0 \delta(p_0+q_0+r_o-k_0)\Xi= -\pi^2.
\end{equation}

 Thus this approach  to the integrals (1.1) dealt with Feynman integrals,
whereas  the first approach worked within the Hamiltonian.
Both the above methods require the external gluons to be transverse. In the Hamiltonian approach, this is because $V_1$ and $V_2$ are functionals 
of the quantum field $\hat{\textbf{A}}^a(t,\textbf{X})$ which is transverse in the Coulomb gauge, that is $\nabla .\hat{\textbf{A}}^a=0$ (the superfix $a$
is for colour and we use hat  to denote quantum fields). In the Feynman integral approach, the external gluons may be taken to be an external classical field $\textbf{A}^a$,
but the derivation of (1.2) \cite{doust} required that this field also must be restricted to be transverse. But the effective action $\Gamma(\textbf{A}^a,\textbf{E}^a, A_0^a)$ is formally defined for general $\textbf{A}^a$, not just transverse. So the question arises whether the divergent integrals (1.1)
in $\Gamma$ can be made convergent as in (1.2) or not? In Sections 3 and 4, we answer this question by `yes', for the simplest example, that is the two-point contribution to $\Gamma$.  

If $\Gamma$ exists in the Coulomb gauge, then we may formally derive BRST identities  (perhaps better called Ward identities, since there are no ghosts) involving it. These would contain terms like (we use $i$ as a spatial index)
\begin{equation}
\frac{\partial}{\partial X_i} \frac{\delta\Gamma}{\delta A_i^a(\textbf{X},t)},
\end{equation}
that is longitudinal part of Green's functions. So the second question is: can we, within the Coulomb gauge, derive $\Gamma$ which would be needed
to verify the identities containing (1.3)? We comment on this in Section 7.

There is a class of gauges, involving a parameter  we call $\theta$, which interpolate between the Coulomb gauge and a covariant gauge,
approaching the former as $\theta\rightarrow 0$. These gauges have no merit in themselves, but they can be used temporarily to regulate integrals
like (1.1). For $\theta\neq 0$, $\Gamma$ is well-defined, and the BRST identities   must be obeyed. We show the limit
\begin{equation}
\lim_{\theta\rightarrow 0}\Gamma(\theta)
\end{equation}
exists graph-by-graph. However, for individual Feynman graphs, the contributions to the effective action (1.4) 
got by this limit cannot be calculated within the Coulomb gauge,
so it seems not to be a useful object.

In the next section, we review the phase space formalism and the interpolating gauge.

\section{The Coulomb gauge and the interpolating gauge}

The phase space formalism was reviewed in, for instance, \cite{AAJCTrenormalization}. The Lagrangian is

\begin{equation}
      -(1/4)F^a_{ij}F^a_{ij}-(1/2)E_i^aE_i^a+E_i^aF^a_{0i}+(\partial_i c^{*a}+u_i^a)(\partial_i c^a+gf^{abc}A_i^bc^c) $$

     $$ +u_0^a[\partial_0c^a+gf^{abc}A_0^b c^c]+gf^{abc}v_i^aE_i^bc^c-(1/2)gf^{abc}K^ac^bc^c.
\end{equation}

Here ($\mu=0,i$),
\begin{equation}
F^a_{\mu\nu}\equiv\partial_{\mu}A^a_{\nu}-\partial_{\nu}A^a_{\mu}+gf^{abc}A^b_{\mu}A^c_{\nu},
\end{equation}
$c,c^*$ are ghost and anti-ghost, and $u_i,u_0,v_i$ are sources used to implement the BRST identities. To the above Lagrangian,
a gauge fixing term is added. We write this for the interpolating gauge, which we choose to be (with a parameter $\theta$)
\begin{equation}
-\frac{1}{2\theta^2}[\partial_iA^a_i-\theta^2\partial_0A^a_0]^2
\end{equation}
where the Coulomb gauge is obtained from  the limit $\theta \rightarrow 0$ (the Feynman gauge is given by
$\theta=1$).
The Feynman rules in the interpolating gauge  and our graphical notation are explained in Fig.1 (neglecting $\theta$ compared to 1).
Dashed lines represent the spatial components of the propagator, dotted lines the time component.
Half continuous lines denote momenta or energies. Doubled lines represent ghosts. In the Coulomb gauge
($\theta =0$) ghost loops just cancel closed loops of Coulomb lines, but in the flow gauge
this is not the case, because the ghosts have a coupling to Coulomb lines.

For our study of the Coulomb gauge, we take the simplest example, the spatial  2-point part of the effective action at 2-loop order, that is at $O(g^4)$, for which we use the notation $\delta^{ab}\Pi_{ij}$ (the colour dependence being trivial).

\begin{figure}[h]
    \centering
    \includegraphics[width=0.8\textwidth]{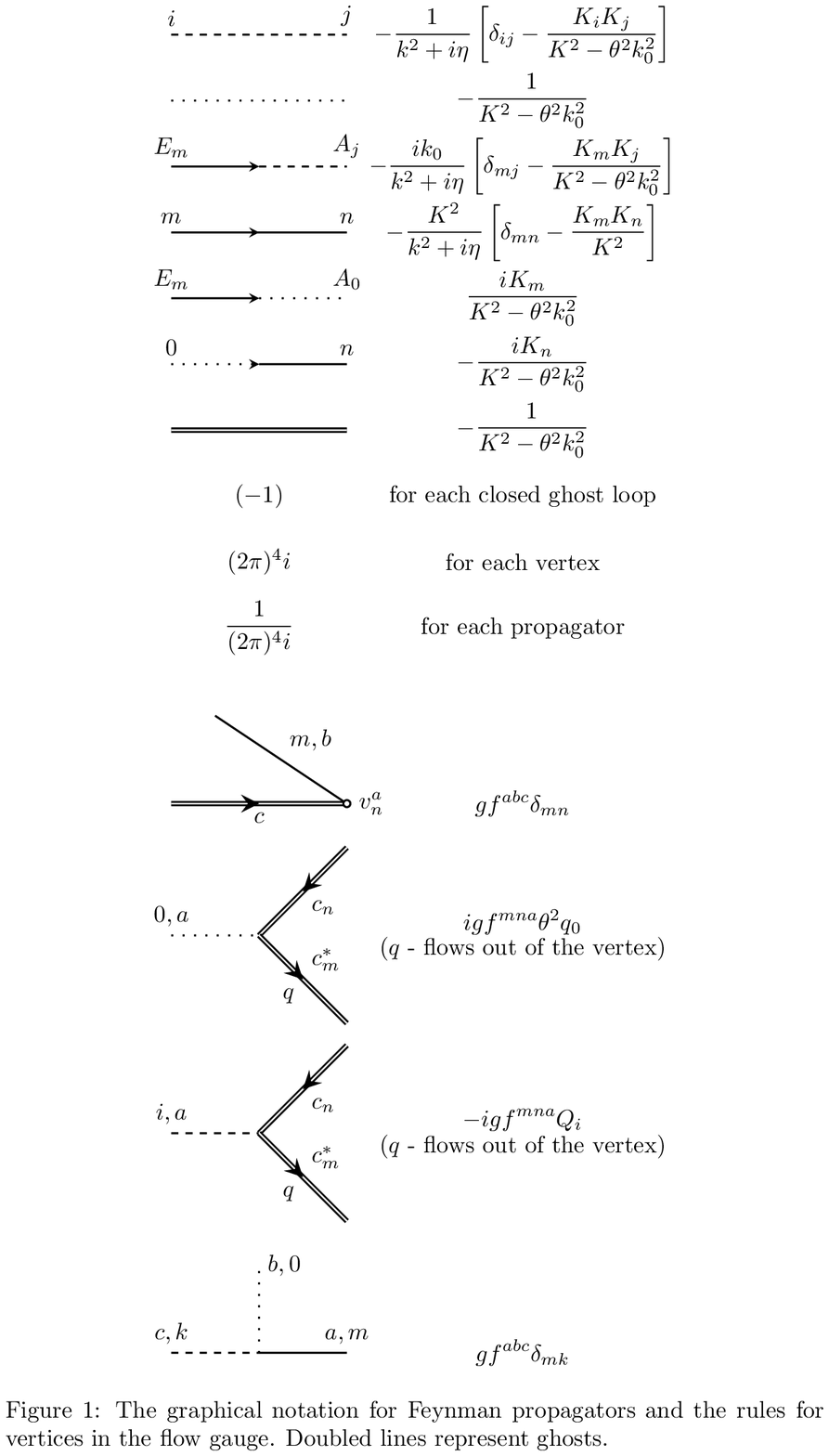}
\end{figure}

\section{The non-convergent, 2-loop integrals}
A Feynman integral, typical of the type we shall study coming from Fig.2A, is (in the flow gauge)
\begin{equation}
\int d^n\textbf{P}d^n\textbf{Q}J(\textbf{P},\textbf{Q},\textbf{K},k_0)
\end{equation}
where
\begin{equation}
J=\lim_{\theta\rightarrow 0}\int dp_0dq_0 \frac{p_0}{p_0^2-{P}^2}\frac{q_0}{q_0^2-{Q}^2}\frac{1}{\bar{P}^2\bar{Q}^2\bar{P'}^2\bar{Q'}^2\bar{R}^2}
\end{equation}
where we use the following notation
\begin{equation}
p'=p-k,\\ q'=q-k,\\\ P=\left|\textbf{P}\right|, \\\ \bar{P}^2=P^2-\theta^2 p_0^2, \\\ r=-p'-q
\end{equation}
etc, and  a Feynman $-i\epsilon$ attached to each $\textbf{P}^2$ etc is understood.

We are not concerned with the spatial integrals in (3.1), which we assume to be controlled by dimensional regularization,
with the spatial dimension $n$ suitably chosen. For the energy integrals, we first  change to the variables
\begin{equation}
\hat{p}_0 = \theta p_0, \\ \hat{q}_0=\theta q_0,
\end{equation}
and then, in the limit, neglect $\theta^2 P^2$, $\theta^2Q^2$, $\theta^2R^2$ and $\theta^2k_0^2$ compared to $\hat{p}_0^2$ etc, so that $\theta$ disappears from the integrand.
 The energy integrals over $\hat{p_0}, \hat{q}_0$ are then done by completing the contours of integration, and give
\begin{equation}
J= -\frac{\pi^2}{(P^2-P'^2)(Q^2-Q'^2)R^2}L
\end{equation}
where
\begin{equation}
L=\frac{P+Q+2R}{PQ(P+Q+R)(P+R)(Q+R)} -(P\rightarrow P')-(Q\rightarrow Q') +( P,Q\rightarrow P',Q')
\end{equation}
Note that the $P^2$ and $Q^2$ in the first two denominators in (3.2) are irrelevant in the limit, and
these denominators might as well be replaced by
\begin{equation}
\frac{p_0}{p_0^2-i\epsilon}\frac{q_0}{q_0^2-i\epsilon}.
\end{equation}
Similarly.
\begin{equation}
\frac{p_0-k_0}{(p-k)^2}\frac{q_0}{q^2}
\end{equation}
can  be replaced by (3.7) with neglect of $O(\theta k_0)$ and $O(\theta^2 P^2)$

Although the result (3.6) is independent of $\theta$, it cannot be obtained by taking the limit $\theta \rightarrow 0$ in the integrand in (3.2)
because the resulting integral would not be convergent. Thus (3.6) is not derivable in the Coulomb gauge. The question is whether there
are combinations of Feynman graphs giving (1.2), which are convergent without the help of the flow gauge.

It is perhaps puzzling that the sub-graphs of Fig.2, taken on there own, would normally be given the value zero
 in the Coulomb gauge. In the flow gauge, the right-hand triangle sub-graph integral gives something of the form
\begin{equation}
\theta p_0 f(\textbf{K},\textbf{P}, \theta p_0, \theta k_0)+O(\theta P, \theta K)
\end{equation}
This does indeed tend to zero as $\theta \rightarrow 0$ for fixed $p_0$; but when this sub-graph is part of the
complete 2-loop graph, $p_0$ is of order $1/\theta$. Thus the value of the sub-graph in the Coulomb gauge (to wit, zero)
cannot be used in the 2-loop graph.

\begin{figure}[h]
    \centering
    \includegraphics[width=0.8\textwidth]{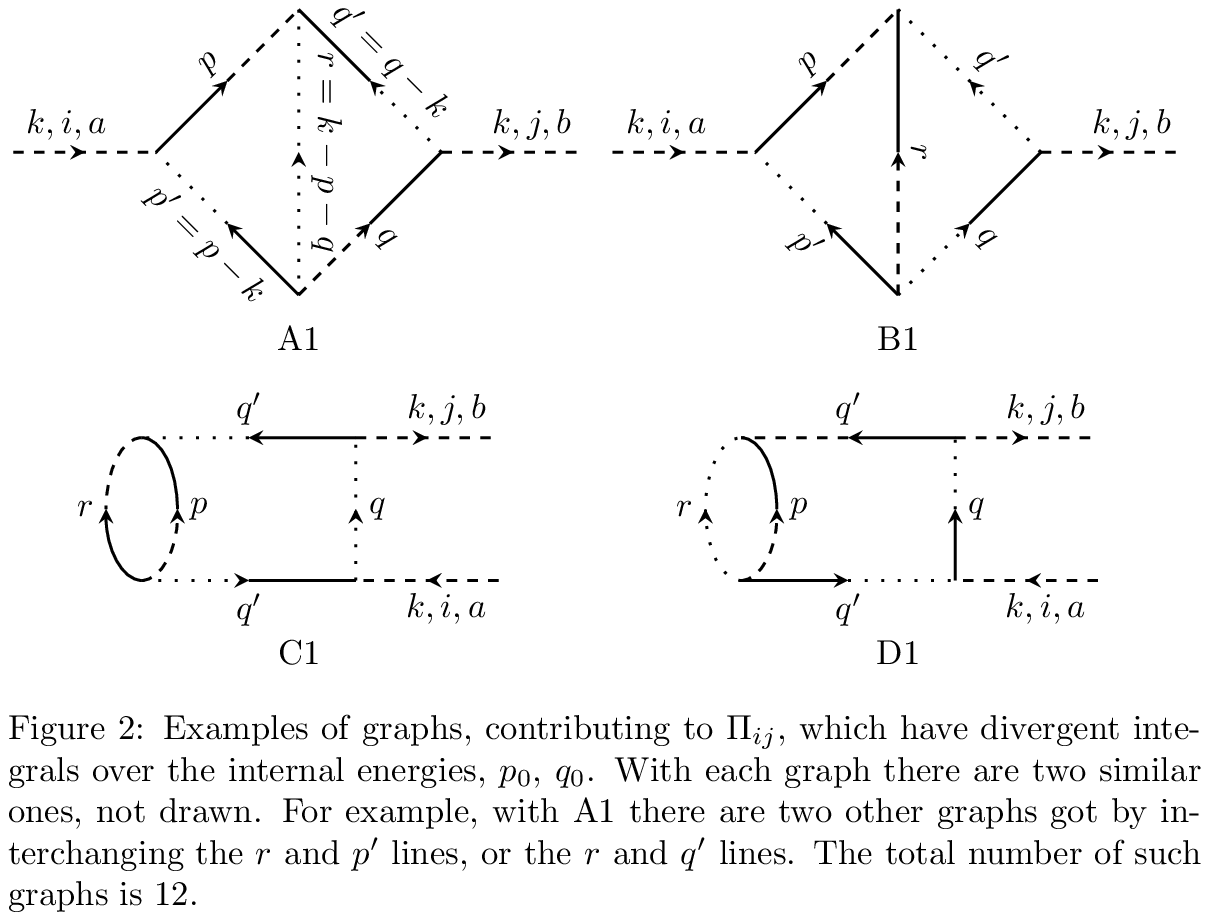}
\end{figure}

\section{Energy integral convergent sets of graphs}

In this section we demonstrate our main result: that, for the 2-gluon function at 2-loop order, Feynman graphs containing the non-convergent
energy integrals (1.1) can be combined
to give the convergent energy integral (1.2). This was proved (as part of a more general theorem) in \cite{doust}, for the special case
of transverse external gluons, that is to say
\begin{equation}
e_ie'_j \Pi_{ij}
\end{equation}
where $\Pi_{ij}$ is the two-gluon part of the effective action (defined in equation (7.3) below), and $\textbf{K}.\textbf{e}=\textbf{K}.\textbf{e}'=0$ ($\textbf{K}$ being the external spatial momentum). We show that the same holds
for general $\Pi_{ij}$. This is not at all obvious, since the proof in \cite{doust} makes frequent use of equations like
$\textbf{P}'.\textbf{e}\equiv (\textbf{P}-\textbf{K}).\textbf{e} =\textbf{P}.\textbf{e}$.

We examine one-by-one  the possible forms of the denominators in the flow gauge, and find that in each case graphs combine to give
the integral (1.2).

We begin with terms containing the 5-factor denominator
\begin{equation}
\frac{1}{\bar{P^2}\bar{Q^2}\bar{R^2}\bar{P'^2}\bar{Q'^2}}
\end{equation}
These come from graphs like Fig.2A,B and Fig.3GA,GB. The factors multiplying (4.2) are
\begin {equation}
(c_n\Xi/8)[3R^4-2R^2(P^2+Q^2+P'^2+Q'^2)-(P^2-Q'^2)(Q^2-P'^2)](P_iQ_j+P'_iQ'_j)$$

$$+(c_n\Xi/8)[R^2(P^2+Q^2-P'^2-Q'^2)](P_iQ_j-P'_iQ'_j)$$

$$-(c_n\Xi/8)[R^2(P^2+Q^2+P'^2+Q'^2)+2(P^2-Q'^2)(Q^2-P'^2)](P_iQ'_j+P'_iQ_j)$$

$$+(c_n\Xi/8)[R^2(Q^2+P'^2-P^2-Q'^2)](P_iQ'_j-P'_iQ_j),
\end{equation}
where $c_n$ is short for
\begin{equation}
c_n=g^4\frac{C_G^2}{2(2\pi)^{n+1}}
\end{equation}
$C_G$ being the colour group Casimir.
In each term in (4.3), the combination $\Xi$ defined in (1.2) appears. The first term in (1.2) comes from graphs like Fig.2A
and like Fig.3GA, the second term from graphs like Fig.2B and Fig.3GB, and the third term from graphs obtained from the
latter reflection in a vertical axis.

In the transverse case (4.1), treated in \cite{doust}, the second and fourth lines of (4.3) are zero, and the first and third lines combine
giving an over-all factor of 3.

Next we come to terms containing the denominator
\begin{equation}
\frac{1}{\bar{Q^2}(\bar{Q'^2})^2\bar{P^2}\bar{R^2}}.
\end{equation} These come from graphs like Fig.2C,D and Fig.3GC,GD.
We find the following coefficient of (4.5):
\begin{equation}
2c_n\Xi [P^2R^2-(\textbf{P}.\textbf{R})^2](Q'_iQ'_j+Q_iQ'_j+Q'_iQ_j)
\end{equation}
Again the combination (1.2) appears. These are the only terms with five factors in the denominator.

Next we study terms with four denominators. One such is
\begin{equation}
\frac{1}{\bar{R^2}\bar{P^2}\bar{P'^2}\bar{Q^2}}
\end{equation}
Terms with
\begin{equation}
\frac{1}{\bar{R^2}\bar{P^2}\bar{Q^2}\bar{P'^2}}, \\\ \frac{1}{\bar{R^2}\bar{P'^2}\bar{Q'^2}\bar{P^2}}, \\\ \frac{1}{\bar{R^2}\bar{P'^2}\bar{Q'^2}\bar{Q^2}}.
\end{equation} can be brought into the form of (4.7) by suitable changes of variables, so we need not consider these separately.
The factor multiplying (4.7) is
\begin{equation}
c_n\Xi (R^2Q_j-\textbf{R.Q}R_j)(6P_i-2K_i),
\end{equation}
with the factor $\Xi$ defined in (1.2) appearing. In the transverse case, (4.1), the $K_i$ term in (4.9) disappears.
(Although (4.9) is not symmetric in $i,j$, the integral of (4.7)$\times$(4.9) must be a linear combination of the symmetric tensors $\delta_{ij}$ and $K_iK_j$.)

Another possible term with four denominators, coming from graphs like B and BG in Figs.2,4,  is
\begin{equation}
\frac{c_n}{8}\int d^nPd^nQ\int dp_0dq_0\frac{p_0r_0}{p_0^2r_0^2}\frac{1}{\bar{P^2}\bar{P'^2}\bar{Q^2}\bar{Q'^2}}H_{ij}(\textbf{P},\textbf{Q})
\end{equation}
where
\begin{equation}
H_{ij}=[3K^2(P+P')_i(Q+Q')_j-(\textbf{P}+\textbf{P}').(\textbf{Q}+\textbf{Q}')K_iK_j$$
$$-2\textbf{P}.(\textbf{Q}+\textbf{Q}')(P+P')_iK_j-2\textbf{P}.(\textbf{P}+\textbf{P}')K_i(Q+Q')_j].
\end{equation}
The $p_0,q_0$ integrals in (4.10) may be done in a similar manner to those in (3.2), and in the limit
$\theta k_0\rightarrow 0$ the result is symmetric in $P,P'$ and in $Q,Q'$. Then, when $H_{ij}$ is inserted from
(4.11) the spatial integral gives zero because of symmetry under $\textbf{Q}\leftrightarrow -\textbf{Q}'$.
Thus the contribution from (4.10) vanishes as $\theta \rightarrow 0$. The same is true for a similar
contribution from graphs obtained from B and BG by interchanging left and right.

There is also the  four-denominator term
\begin{equation}
\frac{1}{\bar{P^2}\bar{Q^2}(\bar{Q'^2})^2}.
\end{equation}
This receives contributions form the C, D  and GC, GD graphs (Figs.3,4), but they cancel in pairs, the C graphs with the GC, and the D graphs with
the GD graphs.

Finally, we deal with terms with three denominators There are four independent possible terms
\begin{equation}
\frac{1}{\bar{P^2}\bar{Q^2}\bar{R^2}},\\\ \frac{1}{\bar{P^2}\bar{Q^2}\bar{P'^2}}, \\\ \frac{1}{\bar{R^2}\bar{Q^2}\bar{Q'^2}},\\\  \frac{1}{\bar{Q^2}(\bar{Q'^2})^2}.
\end{equation}
(Other terms are related to these by changes of variables.) For each of these terms, the multiplying factors turn out to be zero
as a result of cancellations between A, B, C, D graphs and the corresponding GA, GB, GC, GD graphs, in Fig.2 and Fig.3 respectively.

\begin{figure}[h]
    \centering
    \includegraphics[width=0.8\textwidth]{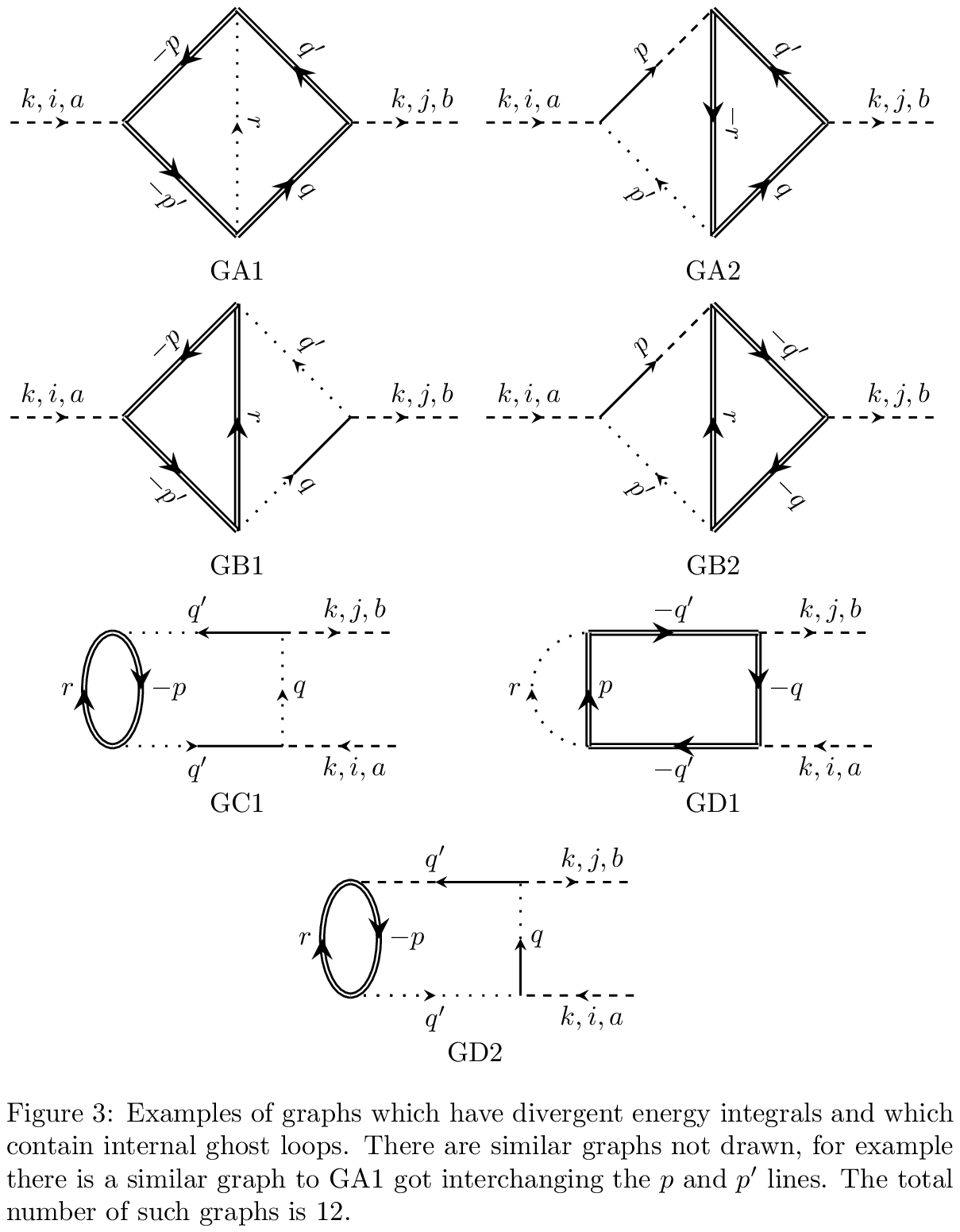}
\end{figure}

\section{The limit $\theta \rightarrow 0$}

Having combined graphs to get the combination $\Xi$ (defined in (1.2)) in (4.3), (4.6) and (4.9), we may let $\theta \rightarrow 0$
in the corresponding denominators (4.2), (4.5) and (4.7), and carry out the energy integrations with the help of (1.3), thus replacing $\Xi$ by $-\pi^2$. Then in the denominators we can replace $\bar{P^2}$ by $P^2$ etc, and cancellations take place between the numerators 
 and the denominators. Many terms such as
\begin{equation}
\int d^nPd^nQ\frac{1}{P^2P'^2(Q'^2)^2},\\\ \int d^nPd^nQ \frac{1}{P^2Q'^2R^2}
\end{equation}
vanish in dimensional regularization (the second is independent of $\textbf{K}$ and the first factorizes with the $\textbf{Q}$-integration independent of $\textbf{K}$). 

What remains can be brought into one of  two forms. The first is
\begin{equation}
-\pi^2c_n\int d^nPd^nQd^nR \delta^n(\textbf{P}+\textbf{Q}+\textbf{R}-\textbf{K})\frac{(3P_iP_j/4)}{P^2Q^2R^2}
\end{equation}

The integrals in (5.2) can be evaluated in configuration space with the aid of the Fourier transform

\begin{equation}
\int d^nP (\textbf{P}^2)^{-m/2}\exp(i\textbf{P}.\textbf{X})=f(n,m)(\textbf{X}^2)^{(m-n)/2}
\end{equation}
where
\begin{equation}
f(n,m)=\frac{\Gamma (\frac{n-m}{2})}{\Gamma (\frac{m}{2})}2^{n-m}\pi^{n/2}.
\end{equation}
The result is that (5.2) is equal to
\begin{equation}
-\pi^2c_n(2\pi)^{-n}(2-n)^2(n-1)[f(n,2)]^3f(n,3n-2)(\textbf{K}^2)^{n-3}[-3\delta_{ij}K^2+3nK_iK_j]
\end{equation}

The pole at $n=3$ is
\begin{equation}
\frac{ c_3\pi^4}{10}\frac{1}{n-3}[\delta_{ij}K^2-3K_iK_j]
\end{equation}
($c_n$ is defined in equation (4.4)).

The second possibility involves the factorisable denominator
\begin{equation}
\frac{1}{P^2P'^2Q^2Q'^2}.
\end{equation}
As explained above, (4.10) does not contribute,
but there is a contribution to  (5.7) coming from (4.2) with the first ($R^4$) term in (4.3). This is
\begin{equation}
-\frac{3c_n\pi^2}{4}\int d^nPd^nQ\frac{R^2P_iQ_j}{P^2P'^2Q^2Q'^2}.
\end{equation}
Writing $R^2=P^2+Q'^2+2\textbf{P}.\textbf{Q}'$ and ignoring integrals independent of $\textbf{K}$, (5.8) factorizes into two integrals:
\begin{equation}
-\frac{3c_n\pi^2}{2}S_{il}(P)S'_{jl}(Q)
\end{equation}
where
\begin{equation}
S_{il}\equiv \int d^nP \frac{P_iP_l}{P^2P'^2},\\\\\\\  S'_{jl}\equiv \int d^nP \frac{P_iP'_l}{P^2P'^2}.
\end{equation}
Then
\begin{equation}
S'_{il}=S_{il}-\frac{1}{2}K_iK_lS,\\\\\ S_{ii}=0,\\\\\ K_iK_lS_{il}=\frac{K^4}{4}S,
\end{equation}
where
\begin{equation}
S(P) \equiv \int d^nP \frac{1}{P^2P'^2}=(2\pi)^{-n}[f(n,2)]^2f(n,2n-4)(K^2)^{\frac{n}{2}-2},
\end{equation}
where $f(n,m)$ is defined in (5.4).
From these equations, we find
\begin{equation}
S_{il}=-\frac{1}{4(n-1)}[K^2\delta_{il}-nK_iK_l]S,$$
$$ S'{il}=-\frac{1}{4(n-1)}[K^2\delta_{il}+(n-2)K_jK_l]S.
\end{equation}
From these equations we deduce that (5.8) gives
\begin{equation}
-\frac{3c_n\pi^2}{2}\frac{1}{(2\pi)^n}[f(n,2)]^4[f(n,2n-4)]^2\frac{1}{16(n-1)^2}[K^2\delta_{ij}-(n^2-2n+2)K_iK_j](K^2)^{n-3}.
\end{equation}
At $n=3$ this has the finite value
\begin{equation}
-6c_3\left(\frac{\pi}{2}\right)^8[K^2\delta_{ij}-5K_iK_j].
\end{equation}

In the transverse case, (4.1), the results (5.2), (5.14) correspond to the function $V_1+V_2$ in the notation of
\cite{christlee} and \cite{doust}.
\section{Other instantaneous graphs}
The graphs in Figs.2,3,4 individually have energy divergences. The convergent combinations, involving $\Xi$ in (1.2) have also the property that they are independent of the external energy $k_0$, that is in configuration space
they are instantaneous. But these are not the only graphs with the latter property. An example is
shown in Fig.5 Here, the energy integrals are 
\begin{equation}
\int dp_0dr_0 \frac{P^2}{p^2}\frac{1}{r^2}=-\pi^2 \frac{\left|\textbf{P}\right|}{\left|\textbf{R}\right|}.
\end{equation}
The expressions appearing in the integrands in (5.2) and (5.9) are rational functions of the vectors $\textbf{P},\textbf{Q}. \textbf{K}$, but for Fig.5, equation (6.1) introduces non-rational dependence;so we expect the values
of these integrals to be quite different from (5.5) and (5.14).

\begin{figure}[h]
    \centering
    \includegraphics[width=0.8\textwidth]{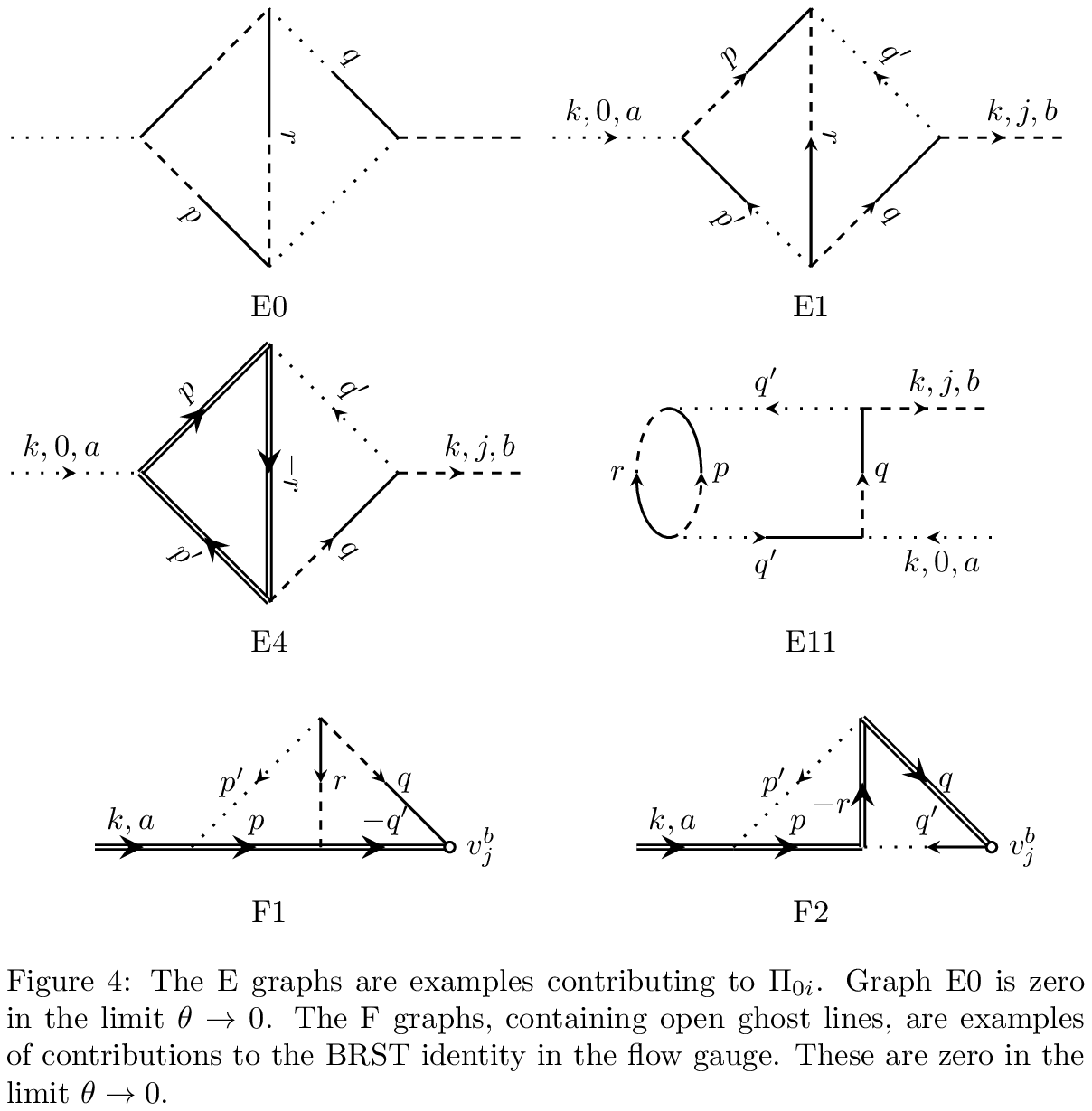}
\end{figure}

\begin{figure}[h]
    \centering
    \includegraphics[width=0.8\textwidth]{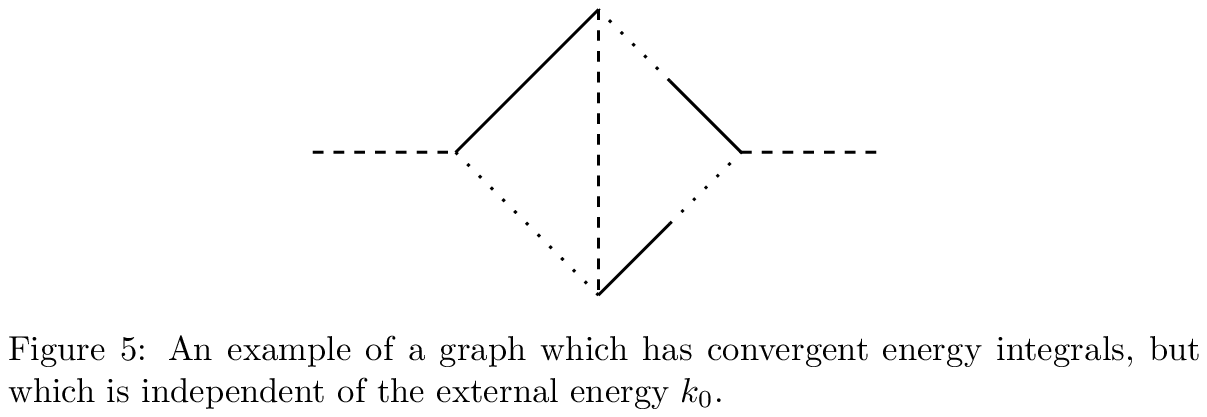}
\end{figure}

\section{The BRST identities}
BRST identities  connect $\Pi_{ij}$ with $\Pi_{0j}$. The latter, with  rather trivial exceptions,  has no energy divergent graphs, and so the flow gauge is not needed to control the integrals.
An example of an exception is graph E0 in Fig.4 In the Coulomb gauge, this contains 
energy integral
\begin{equation}
\int dp_0dr_0 \frac{p_0}{p^2}\frac{p_0-k_0}{(p_0-k_0)^2}\frac{r_0}{r^2},
\end{equation}
where the $r_0$ integral is divergent; so in order to make this unambiguous we must approach
it as the limit from the flow gauge. This introduces factors like
\begin{equation}
\frac{1}{(\textbf{R}+\textbf{P})^2-\theta^2(r_0+p_0)^2}
\end{equation}
and the limit as $\theta \rightarrow 0$ is zero ($r_0$ is of order $1/\theta$, but $p_0$ is not).
Thus the only energy divergent graphs in $\Pi_{0j}$ can be taken to be zero.

There can be no doubt about the validity of the BRST identities in the flow gauge. And the limits
as $\theta\rightarrow 0$ of both $\Pi_{0j}$ and $\Pi_{ij}$ exist, the latter in the sense that graphs can be combined,
as we have shown in section 4, to give the convergent combination $\Xi$ in (1.2). In this sense,
we expect the BRST identity to hold in the Coulomb gauge.

%The BRST identities \cite{itzykson} are (it is to be understood that all the field afrguments of $\Gamma_4$ are to $be put to zero
%after the functional derivatives have been taken)
%\begin{equation}
%\frac{\partial}{\partial y_{\mu}}\frac{\delta^2\Gamma_4}{\delta A^b_{\mu}(y)\delta %A^a_i(x)}=-T_{ij}(x)\frac{\delta^2\Gamma_4}{\delta u^a_j(x) \delta c^b(y)}+\frac{\partial}{\partial% x_0}{\frac{\delta^2\Gamma_4}{\delta v_i^a(x)\delta c^b(y)},
%\end{equation}
%\begin{equation}
%\frac{\partial}{\partial y_{\mu}}\frac{\delta^2\Gamma_4}{\delta A^b_{\mu}(y)\delta %A^a_0(x)}=-T_{ij}(x)\frac{\delta^2\Gamma_4}{\delta u^a_i(x) \delta c^b(y)}-\frac{\partial}{\partial% x_j}\frac{\delta^2\Gamma_4}{\delta v^a_j(x)\delta c^b(y)}+\frac{\partial}{\partial x_0}\frac{\delta^2\Gamma_4}{\delta v^a_j \delta c^b(y)},
%\end{equation}
%here the transverse projection operator is
%\begin{equation}
%T_{ij}(x)\equiv \frac{\partial^2}{\partial x_i \partial x_j}-\delta_{ij}\frac{\partial^2}{\partial x_k\partial x_k}
%\end{equation}

%W have omitted terms like
%\begin{equation}
%\int dz \frac{\delta^2\Gamma_2}{\delta A^a_i(x)\delta A^c_j(z)}\frac{\delta^2 \Gamma_2}{\delta u_j^c(z) \delta c^b(y)}
%\end{equation}
%because they lead to irrelevant integrals of the form of a product
%\begin{equation}
%\int dp f(p,k) \int dq h(q,k).
%\end{equation}
In order to write down the relevant BRST identities in the flow gauge, we first
 define Fourier transforms, like
\begin{equation}
\frac{\delta^2\Gamma_4}{\delta A_i^a(x) \delta A_j^b(y)}\equiv\int dk \exp[-ik_0(x-y)+i\textbf{K}.(\textbf{X}-\textbf{Y}]\delta^{ab}\Pi_{ij}(k)
\end{equation}
where $\Gamma_4$ is the effective action to order $g^4$ and all its field arguments are to be put zero after
the functional derivatives are taken. $\Pi_{0j}$ is defined similarly, and $\Pi^{(u)}_i, \Pi^{(v)}_i$ are respectively the Fourier transforms of
\begin{equation}
\frac{\delta^2 \Gamma_4}{\delta u_i(x)\delta c(y)}, \\\\\\\\ \frac{\delta^2 \Gamma_4}{\delta v_i(x) \delta c(y)}.
\end{equation}
With this notation, the BRST identity  is
\begin{equation}
 K_i \Pi_{ij}=-k_0 \Pi_{0j}-T_{jl}(K) \Pi^{(u)}_j+k_0 \Pi^{(v)}_j,
\end{equation}
where
\begin{equation}
T_{jl}(K)\equiv K^2\delta_{jl}-K_jK_l
\end{equation}
By rotational invariance, $\Pi^{(u)}_j$ is proportional to $K_j$, so the second term on the right of (7.5) is zero.

The term $\Pi^{(v)}_j$ in (7.5) is needed in the flow gauge, and examples of contributing graphs labeled F in Fig.4. But these terms all vanish in the limit $\theta \rightarrow 0$ because of the factor $\theta$
in the coupling of Coulomb lines to ghosts (see Fig.1). So, in the Coulomb gauge, only the first term on the right of (7.5) survives, and the identity reduces to a simple Ward identity.

 Thus there must be contributions from
$\Pi_{0j}$ in (7.5) which balance the $k_0$-independent contributions (5.2) and (5.9) to $\Pi_{ij}$.
Two examples, E1 and E11, of graphs which might be relevant are shown in Fig.4. For example,
E1 contains the energy integral
\begin{equation}
\int dp_0dq_0 dr_0 \delta(p_0+q_0+r_0-k_0)\frac{p_0q_0r_0}{p^2q^2r^2}=-\pi^2\frac{k_0}{k_0^2-(P+Q+R)^2}
\end{equation}
which makes a non-zero contribution to (7.5) in the limit $k_0 \rightarrow \infty$.
However, we find that graphs E1 and E11 by themselves are not sufficient to balance the $k_0$-independent terms
in the Ward identity (7.5).

\section{Summary}
We have studied an example of 2-loop Feynman integrals in Coulomb gauge QCD which individually have energy divergences, and shown how they can be combined to give convergence. This has previouly been done only for
transverse external fields. In particular, the all-orders proof of \cite{doust} made extensive use of transverality. Our result suggests that it may be possible to modify the argument in \cite{doust} so that it  makes no
 use of transversality.
\begin{acknowledgment}
\begin{flushleft}

\end{flushleft}

We are most grateful to Bruno Klajn for preparing the figures.
\end{acknowledgment}

%\begin{figure}[htbp]
	%	\includegraphics[width=15cm,height=5cm]{Fig5.pdf}
	%\label{fig:Fig5}
%\end{figure}

\end{document}